\documentclass[sigconf,screen]{acmart}

\newcommand{\paperTitle}{Mutation Testing of Simulink Cyber-Physical System Models: Challenges and Solutions in Practice}

\usepackage{hyperref} 
\usepackage{xspace}
\usepackage{color}
\usepackage{xcolor}
\usepackage{fancybox}
\usepackage{fancyhdr}
\usepackage{ifthen} 
\usepackage{paralist} 
\usepackage{graphicx}
\usepackage{booktabs}
\usepackage{amsmath}
\usepackage{tcolorbox}
\usepackage{colortbl}
\usepackage{xcolor}
\usepackage{float}
\usepackage[T1]{fontenc}

\usepackage{csquotes}

\definecolor{GraphFrameBG}{RGB}{254,250,240}
\newenvironment{qblock}{%
\begin{displayquote}\begin{tcolorbox}[%
    width=\linewidth,
    size=small,
    colframe=black,
    colback=GraphFrameBG,
    nofloat,
    fontupper=\itshape
]}{\end{tcolorbox}\end{displayquote}}

\newboolean{acmtemplate}
\setboolean{acmtemplate}{true} 
\newboolean{anonymous}
\setboolean{anonymous}{false}
\newboolean{includecomments}
\setboolean{includecomments}{true}

\ifthenelse{\boolean{anonymous}} {

}{

}

\newcommand{\secref}[1]{Section~\ref{#1}\xspace}
 
\newcommand{\tabref}[1]{Table~\ref{#1}\xspace}

\ifthenelse{\boolean{includecomments}} {
    \newcommand{\nb}[3]{
    	{\colorbox{#2}{\bfseries\sffamily\scriptsize\textcolor{white}{#1}}}
    	{\textcolor{#2}{\sf\small$\blacktriangleright$\textit{#3}$\blacktriangleleft$}}}
} {
    \newcommand{\nb}[3]{}
}

\newcommand{\mutslx}{\textsf{MUT4SLX}\xspace}
\newcommand{\duco}{\textsf{DUCO}\xspace}

\newcommand*{\RQOne} [1] {What are the challenges of applying mutation to Simulink and Stateflow models in practice?}
\newcommand*{\RQTwo} [1] {What are the benefits of mutation testing over and beyond existing coverage metrics?}

\newcommand*{\RQTwoOne} [1] {How strongly do traditional Simulink/Stateflow coverage metrics correlate with mutation coverage?}

\newcommand*{\RQTwoTwo} [1] {Is the association consistent across different model types?}

\newcommand*{\RQThree} [1] {What are the possible solutions to the encountered challenges?}

{\ttfamily \hyphenchar\the\font=`\-}

\hypersetup{
    colorlinks,%
    citecolor=black,%
    filecolor=black,%
    linkcolor=black,%
    urlcolor=gray,
    linktocpage=true,
    bookmarksopen=true,
    pdfpagemode=UseOutlines,
}

\setcopyright{none}
\renewcommand\footnotetextcopyrightpermission[1]{}
\begin{document}
\title{\paperTitle} 
\titlenote{Accepted for publication at the 41st IEEE/ACM International Conference on Automated Software Engineering (ASE 2026), Munich, Germany. Version of Record: \url{https://doi.org/10.1145/3832783.3834529}}

        \author{Murat Kavak}
        \affiliation{%
            \institution{Universiteit~Antwerpen}
            \country{Belgium}
        }
        \affiliation{%
            \institution{Flanders~Make~VZW}
            \country{Belgium}
        }

        \author{Onur Kilincceker}
        \affiliation{%
            \institution{Universiteit~Antwerpen}
            \country{Belgium}
        }
        \affiliation{%
            \institution{Flanders~Make~VZW}
            \country{Belgium}
        }

        \author{Serge Demeyer}
        \affiliation{%
            \institution{Universiteit~Antwerpen}
            \country{Belgium}
        }
        \affiliation{%
            \institution{Flanders~Make~VZW}
            \country{Belgium}
        }

        \author{Kevin Vandenbroucke}
        \affiliation{%
            \institution{DUCO (NV)}
            \country{Belgium}
        }

        \author{Halim Abdurrahman Ceylan}
        \affiliation{%
            \institution{Department of International Computer, Graduate School of Natural and Applied Sciences, Ege University, Izmir}
            \country{T{\"u}rkiye}
        }

        \renewcommand{\shortauthors}{Kavak et al.}

\begin{abstract}
Several experience reports illustrate that mutation testing is capable of supporting a ``shift-left'' testing strategy, especially in industries where late bug discovery incurs very high costs and risks.
In the context of cyber-physical systems, a ``shift left''  implies that engineers need to test the design models used to simulate, prototype, and analyse the feasibility of the system under design.
In this paper, we analyse the challenges we encountered and the lessons we learned when incorporating mutation testing in the context of \duco, a company producing ventilation systems for buildings.
The engineers within \duco have years of experience with model-based engineering centered around Simulink and StateFlow, including unit tests for their models.
During a pilot project with our tool prototype \mutslx we learned that equivalent mutants, requirement traceability, and mutation testing for Stateflow represent particular challenges not yet reported in the academic literature. 
\end{abstract}


\maketitle
\thispagestyle{plain}
\pagestyle{plain}

\section{Introduction}
\label{sec::Intro}

Safety-critical industries (such as automotive and aerospace) mainly rely on model-driven engineering to tackle the complexity and pervasiveness of safety-critical software. 
This reliance is even more apparent in cyber-physical systems, where software interacts with physical components and processes under strict safety, timing, and reliability constraints~\cite{lee2015past}.

Testing cyber-physical systems is very challenging due to their complexity and heterogeneity, characterized by dependencies among software, physical components, and processes. 
Shift-left testing is a software development approach that aims to move testing activities into an early phase of the lifecycle, when dependencies and interactions are simulated using executable models of the system and its environment~\cite{gonzalez2018enabling}.
This approach is coined as model-based shift-left testing~\cite{miller2021four}, which offers several advantages for ensuring the safety and reliability of modern cyber-physical systems. First, the system is modeled at a higher level of abstraction to manage complexity. Second, since it is an early phase, the cost of fixing is 10 to 30 times lower than that of a bug detected in later development stages~\cite{planning2002economic}.

Today, MathWorks Simulink\textregistered and Stateflow\textregistered offer an advanced and widely used platform for simulation and development of cyber-physical systems~\cite{matinnejad2016automated}. Also, it enables the application of model-based shift-left testing principles to cyber-physical systems.
Simulink models allow developers to design systems using block diagrams that transform input signals into output signals, thereby capturing the behavior of complex and dynamic systems.
Stateflow, on the other hand, is more effectively used to model and simulate decision logic using state machines and flowcharts.
The best practice is to use Stateflow for control and decision logic, alongside Simulink, to develop a cyber-physical system.
This way, Simulink's capabilities are extended to event-driven and hybrid forms of embedded control via Stateflow.

Mutation testing is a recommended practice in industrial standards in automotive, industrial robots, and medical devices for functional safety, such as ISO 26262. 
The main purpose of mutation testing is to diagnose and address weaknesses in software test suites, typically expressed in textual programming languages such as C++~\cite{demillo2006hints}.
This method intentionally injects faults (mutants) into the code or model and measures how many are detected by the test suite, producing a mutant score that reflects the suite’s quality~\cite{baker2012empirical}. 
Based on shift-left testing principles, mutation testing at the early design model-level stage offers advantages over code-level testing, since the system is simpler and the detected faults are less costly and easier to fix.
In the literature, it is commonly referred to as model-based mutation testing~\cite{aichernig2015killing, belli2016model}.
To automate this approach for Simulink and Stateflow, SIMULTATE~\cite{pill2016simultate}, FIM~\cite{bartocci2022fim}, \mutslx{}~\cite{ceylan2023mut4slx, nuyens2024mut4slx}, and BERTiMuS~\cite{zhang2025simulink} are proof-of-concept tools, each focusing on different aspects and using different methods.

While these proof-of-concept tool demonstrate the technical feasibility of model-based mutation testing, actual experience reports on how the corresponding tools would work in practice are still lacking.
In this paper, we analyse the challenges we encountered and the lessons we learned when incorporating mutation testing in the context of \duco.
\duco adopted model-based engineering practices as early as 2019. 
At the time of writing, an average project incorporates 20\% Simulink\textregistered and 80\% Stateflow\textregistered models.
Unit testing is heavily promoted within the \duco; by means of the Signal Builder (now Signal Editor)\footnote{\url{https://nl.mathworks.com/help/simulink/ref_obsolete_blocks/signalbuilder.html} (accessed 2026-03-30).} block for test creation in combination with an Assertion block to check whether the actual result matches.
\duco measures the coverage of their unit tests but has noticed that defects still appear later in the life cycle.
In this context, they were eager to validate our tool prototype \mutslx within their workflow.

The remainder of this paper is organised as follows.
\secref{sec::RelatedWork} gives an overview of the state of the art in mutation testing for Simulink and Stateflow models.
\secref{sec::Questions} describes the research questions, which naturally leads to \secref{sec::Results_Discussion} reporting the results. \secref{sec::lessons} presents lessons learned, including practical insights, and concludes the paper in \secref{sec::Conclusion}.


\section{Related Work}
\label{sec::RelatedWork}

Fault injection is used to test the robustness of the system under test, whereas mutation testing is used to measure test effectiveness.
Both methods inject faults in a systematic way, based on well-defined and structured fault patterns.
The fault patterns are (ideally) defined based on specific bug databases to represent realistic errors unintentionally injected by developers (the Competent Programmer hypothesis~\cite{budd1980theoretical}), and these errors may couple and lead to complex failures (the Coupling Effect hypothesis~\cite{budd1980theoretical}).
Yadav et al. presented a comprehensive analysis and comparison of state-of-the-art fault-injection tools for Simulink Cyber Physical System models, including insights and future direction of research in this research area~\cite{yadav2025fault}.

Mutation testing for Simulink models dates back to 2005, when Zhan and Clark introduced three types of mutation operators (add, multiply, and assign) for the signal carried on the input by/with a certain value~\cite{zhan2005search}.
Those operators are primarily used for search-based test generation in combination with mutation testing.
Brillout et al. employed mutation testing again to generate bounded model-checking-based test data for Simulink models~\cite{brillout2009mutation}.
Complementary to the above, there are a few experience reports concerning mutation testing (and the companion technique fault injection) in model-based engineering.
Rana et al. presented the results of the first ISO 26262-compliant verification and validation efforts, combining fault injection and mutation testing for Simulink models~\cite{rana2013early}.
This was a collaborative research between the University of Gothenburg and Volvo Car Corporation in the automotive industry.
They concluded that fault injection and model-level mutation testing (as part of shift-left testing) add value to the development of high-quality, reliable, and dependable software in the automobile industry, help keep product costs lower, and, most importantly, ensure that automobiles are safer than ever before.

Early research on developing a taxonomy of classes for Simulink model mutations proposed 4 categories to analyze simple faults that designers may commit~\cite{binh2012mutation}.
Then, a more extensive set of 8 proposed categories was devised to test Simulink model-clone detectors~\cite{stephan2014towards}.
Table~\ref{tab:simulink-mutation-classes} shows those classes with underlying change logic, including corresponding related work fitting to those classes. 

\begin{table}[htbp]
\centering
\caption{Simulink Mutation Classes~\cite{stephan2014towards}}
\label{tab:simulink-mutation-classes}
\scriptsize
\renewcommand{\arraystretch}{1.1}
\setlength{\tabcolsep}{2.5pt}
\begin{tabular}{|p{1.3cm}|p{2.6cm}|p{2.9cm}|}
\hline
\textbf{Mutation Key} & \textbf{Title} & \textbf{Related Work} \\ \hline
mMLA  & Modification of Layout Attribute & \\ \hline
mRUE  & Reordering Underlying Elements & \cite{pill2016simultate, matinnejad2018test} \\ \hline
mRBL  & Renaming a Block or Line & \\ \hline
mCBV  & Changing a Block's Value & \cite{pill2016simultate, Hanh2016Fitness, arrieta2017towards, nanda2017integrated, matinnejad2018test, bartocci2022fim, ceylan2023mut4slx, nuyens2024mut4slx}\\ \hline
mADBD & Add or Delete Block as Destination & ~\cite{matinnejad2018test}\\ \hline
mADBS & Add or Delete Block as Source & \\ \hline
mCBT  & Changing a Block's Type & ~\cite{bartocci2022fim}\\ \hline
mCSCH & Changing a Subsystem's clone hierarchy & \\ \hline
\end{tabular}
\end{table}

The first prototype tool, SIMULTATE, automates mutation testing for Simulink models~\cite{pill2016simultate}. It relies on a Python interface to automate processes, allowing users to expand with additional mutation operators (initially proposing 4) and to support fault injection (including 5 fault models). Its operators fall into the category of mRUE and mCBV (shown in Table~\ref{tab:simulink-mutation-classes}). Then, Hanh et al. presented a test data generation approach for Simulink models based on mutation analysis, proposing 10 mutation operators that fall into the mCBV category~\cite{Hanh2016Fitness}. Those operators are also implemented in a tool using feature diagrams (inspired by Devroey et al.~\cite{devroey2016featured}) to generate mutants for configurable Simulink models~\cite{arrieta2017towards}. Nanda et al. developed the IMAT tool, which integrates with Simulink Design Verifier (automatic test generator) to connect mutation testing to the model checker~\cite{nanda2017integrated}. Matinnejad et al. presented a test generation algorithm and a test prioritization algorithm for Simulink models, using mutation testing to measure the fault-revealing abilities of the proposed algorithms~\cite{matinnejad2018test}. A comprehensive list of Simulink and Stateflow fault patterns was also presented and compared with both the existing literature on mutation operators and the identified ones by Delphi Automotive.

FIM is a proof-of-concept tool that inherits the mutation operators of the SIMULATE tool and supports ten fault injection operators for signals and five mutation operators for blocks~\cite{bartocci2022fim}. The tool primarily focuses on safety analysis through extensive fault-injection capabilities. FIM is also extended with additional operators and property-based mutation testing to assess a test suite's ability to exercise the software with respect to a given property~\cite{bartocci2023property}.

\mutslx{} is a state-of-the-art mutation testing tool capable of both automated mutant generation and execution~\cite{ceylan2023mut4slx, nuyens2024mut4slx}. It includes a comprehensive set of parameter-based mutation operators—currently 28 for Simulink and 18 for Stateflow—modeled after realistic faults mined from an industrial bug database. It is also a unique tool supporting both requirement traceability and CI/CD pipeline compatibility. Currently, \mutslx{} is at technology readiness level (TRL) 5 and commercial readiness level (CRL) 4. This is why \duco considered only \mutslx{} rather than other tools since \mutslx{} was tailored based on \duco requirements and development pipeline.

Unlike functional operators, Chen et al. introduced 6 high-level timed mutation operators for Stateflow models, considering the importance of real-time behavior during simulation, and employed them to evaluate SimSched, a tool that utilizes model transformation~\cite{chen2024timed}. 

BERTiMuS is a mutant generation approach that uses large language models (based on CodeBERT, a pre-trained model for programming languages) for Simulink models~\cite{zhang2025simulink}. Valle et al. also presented an automated pipeline for Simulink-Stateflow mutant generation and evaluated eight state-of-the-art LLMs with different prompting strategies~\cite{valle2026exploring}.

Unlike others, our work focuses on the challenges encountered and potential solutions in applying mutation testing to Simulink Cyber-Physical System models using \mutslx{} in practice, based on a pilot project with \duco company. 

\section{Driving the Pilot Project}

\subsection{Method}
\label{sec::Method}

The pilot project was conducted within an existing research and development project called efficient testing of control software (EFFECTS) where there was a dedicated work package on mutation testing of Simulink and Stateflow models in collaboration with \duco company. The project lasted between 2021 and 2024. One of the deliverables of the relevant work package was a proof-of-concept tool (called \mutslx{}) and we first released our minimum-viable product (MVP) in 2023 (Q1). Since \mutslx{} was succesfull on providing measurable benefits (as we share in the next section) on DUCO side, we decided to further invest on mutation testing using \mutslx{} and also collected extra feature requests and comments on our proof-of-concept tool after finalizing our project. Then, we regularly share new versions addressing those features and then collect feedback from \duco side. 

\subsection{Quantitative Details}
\label{sec::QuantitativeDetails}

The pilot was carried out on three production models taken from \duco's model repository, referred to
here as \textbf{M1}, \textbf{M2} and \textbf{M3}; their names and internal structure cannot be
disclosed (see \secref{sec::Availability}).
\duco applied \mutslx internally to a considerably larger set of models and reported back a 7.5\%
improvement of their test suites over the first three consecutive trials of the tool.
The challenges and solutions discussed in this paper emerged from that wider internal use and from
continuous discussion with the \duco engineers.
These three models are the ones shared with us for joint analysis and hence the only ones we can report
measurements on.
M1 and M2 are Simulink\textregistered{} industrial control components, a saturating summation stage and
a scaling PID controller. M3 is a Stateflow\textregistered{} controller from the heating domain.
\tabref{tab:industrial_models} summarises their size and the structural coverage their existing test
suites achieve.
The models are exercised exactly as \duco exercises them in daily practice.
A test case is a Signal Builder group that drives one input scenario, and the expected behaviour is
checked by Assertion blocks embedded in the model.
\mutslx therefore counts a mutant as killed whenever running the test suite triggers an assertion.
We ran the complete operator set without any filtering, so the numbers below are raw results.
All measurements were obtained with MATLAB\textregistered{} R2025b on a laptop with an Intel Core
i7-12700H processor (14 cores, 20 threads), 16\,GB of RAM and Windows 11.

\begin{table*}[!t]
\centering
\caption{Characteristics of the three industrial models used in the pilot, and the structural coverage
reached by their existing test suites when all test cases are active.}
\label{tab:industrial_models}
\footnotesize
\renewcommand{\arraystretch}{1.2}
\setlength{\tabcolsep}{9pt}
\begin{tabular}{llrrcrrrrr}
\toprule
& & \multicolumn{3}{c}{\textbf{Component under test}} & \multicolumn{2}{c}{\textbf{Test suite}} & \multicolumn{3}{c}{\textbf{Coverage (\%)}} \\
\cmidrule(lr){3-5}\cmidrule(lr){6-7}\cmidrule(lr){8-10}
\shortstack{\textbf{Model}} & \shortstack{\textbf{Model}\\\textbf{type}} & \shortstack{\textbf{Blocks}} & \shortstack{\textbf{Subsystems}} & \shortstack{\textbf{States /}\\\textbf{Trans.}} & \shortstack{\textbf{Test}\\\textbf{cases}} & \shortstack{\textbf{Input}\\\textbf{signals}} & \shortstack{\textbf{Decision}} & \shortstack{\textbf{Condition}} & \shortstack{\textbf{MC/DC}} \\
\midrule
\textbf{M1} & Simulink & 49 & 2 & N/A & 5 & 4 & 100.0 & 100.0 & N/A \\
\textbf{M2} & Simulink & 255 & 11 & N/A & 21 & 18 & 69.6 & 60.0 & N/A \\
\textbf{M3} & Stateflow & 28 & 3 & 6 / 14 & 1 & 8 & 100.0 & 100.0 & 100.0 \\
\bottomrule
\end{tabular}

\vspace{2pt}
\begin{minipage}{\textwidth}
\centering
\footnotesize
\textit{Abbreviations:} Trans.\ = transitions; MC/DC = modified condition/decision coverage;
N/A = the model contains no element for which the metric is defined.
Coverage is measured on the component under test, not on the surrounding test model.
\end{minipage}
\end{table*}

Over the three models \mutslx generated 459 mutants, of which 250 were killed and 209 survived,
yielding mutation scores of 82.0\% (M1), 41.0\% (M2) and 95.7\% (M3).
M1 reaches 100\% decision and condition coverage, yet 16 of its 89 mutants survive.
The 209 surviving mutants are also the pool from which equivalent mutants have to be identified by
hand, which is the challenge discussed in \secref{rq1_challenges}.
Execution cost is a further practical concern: M2 required 3.6 hours to simulate its 324 mutants
against its 21 test cases.

\begin{table}[!t]
\centering
\caption{Mutation score of each test case in isolation compared to the score of the complete test suite.}
\label{tab:per_testcase}
\footnotesize
\renewcommand{\arraystretch}{1.2}
\setlength{\tabcolsep}{6pt}
\begin{tabular}{lrrrr}
\toprule
& & & \multicolumn{2}{c}{\textbf{Run time}} \\
\cmidrule(lr){4-5}
\textbf{Test case} & \textbf{Killed} & \textbf{Score (\%)} & \textbf{(s)} & \textbf{(min)} \\
\midrule
\multicolumn{5}{l}{\textit{\textbf{M1}} (5 test cases)} \\
\cmidrule(lr){1-5}
\quad TC1 & 40 & 44.9 & 16.9 & 0.3 \\
\quad TC2 & 40 & 44.9 & 16.9 & 0.3 \\
\quad TC3 & 40 & 44.9 & 17.1 & 0.3 \\
\quad TC4 & 41 & 46.1 & 16.9 & 0.3 \\
\quad TC5 & 61 & 68.5 & 17.0 & 0.3 \\
\quad \textbf{All 5 test cases} & \textbf{73} & \textbf{82.0} & \textbf{76.7} & \textbf{1.3} \\
\midrule
\multicolumn{5}{l}{\textit{\textbf{M2}} (21 test cases)} \\
\cmidrule(lr){1-5}
\quad Weakest test case & 81 & 25.0 & 242.9 & 4.0 \\
\quad Median & 124 & 38.3 & 224.8 & 3.7 \\
\quad Strongest test case & 130 & 40.1 & 368.2 & 6.1 \\
\multicolumn{5}{l}{\quad\footnotesize 16 of the 21 test cases score between 37.7 and 38.9} \\
\quad \textbf{All 21 test cases} & \textbf{133} & \textbf{41.0} & \textbf{12823.4} & \textbf{213.7} \\
\midrule
\multicolumn{5}{l}{\textit{\textbf{M3}} (1 test case)} \\
\cmidrule(lr){1-5}
\quad \textbf{TC1 (= all test cases)} & \textbf{44} & \textbf{95.7} & \textbf{24.8} & \textbf{0.4} \\
\bottomrule
\end{tabular}

\vspace{2pt}
\begin{minipage}{\linewidth}
\centering
\footnotesize
\textit{Note:} Test cases are anonymised and ordered by increasing mutation score. The mutant pool is
identical for every row of a model (M1: 89, M2: 324, M3: 46 mutants). For \textbf{M2} the three
summary rows report the minimum, median and maximum of each column over its 21 single-test-case runs.
\end{minipage}
\end{table}

\tabref{tab:per_testcase} contrasts each test case in isolation with the complete suite.
For M1 the individual test cases kill between 44.9\% and 68.5\% of the mutants while the suite as a
whole reaches 82.0\%, so the test cases complement each other.
For M2 the picture is different. Its strongest single test case already reaches 40.1\% and the full
suite of 21 only reaches 41.0\%, with 16 of the 21 test cases scoring between 37.7\% and 38.9\%.
The suite is therefore largely redundant with respect to fault detection, an insight that structural
coverage alone did not expose.
For M3 (the smallest of the three subjects) there is only 1 test cases killing 44 mutants and yielding 95.7\% mutation score. Also note that even there full decision, condition and MC/DC coverage leaves two of its 46 mutants alive.

\subsection{Research Questions}
\label{sec::Questions}

In this paper, we report on a pilot project for a proof-of-concept model-based mutation testing tool named \mutslx within the context of \duco.
\duco has a long history of model-based engineering, including a heavy emphasis on unit tests for its Simulink and Stateflow models.
The pilot study was driven by two primary research questions listed below.

\renewcommand*{\RQOne} [1] {What are the challenges of applying mutation to Simulink and Stateflow models in practice?}
\renewcommand*{\RQTwo} [1] {What are the possible solutions to the encountered challenges?}

\begin{qblock}
\textbf{RQ1: Challenges} \textit{\RQOne}
\end{qblock}

\noindent\textbf{Motivation.} We experienced several challenges when introducing a model-based mutation testing tool within the context of \duco.
While specific to the context of \duco, we believe they are sufficiently generic to warrant attention of the research community.

\noindent\textbf{Approach.} We share the challenges we encountered, considering their importance and severity. We give priority to the most severe challenges (\textsf{major)}, including some relevant minor ones in between.

\begin{qblock}
\textbf{RQ2: Possible Solutions} \textit{\RQTwo}
\end{qblock}

\noindent\textbf{Motivation.}
During the pilot project we addressed these challenges to work within the context of the \duco engineering team.
Here as well our lessons learned may benefit the larger research community.

\noindent\textbf{Approach.} 
Based on the feedback of the \duco engineering team, we reviewed the existing literature relevant to the feedback, highlighted topics and considered possible solutions.
Then, we explored whether it is feasible to incorporate these solutions in the context of Simulink and Stateflow, expanding \mutslx with extra features.
We document our lessons learned.


\section{Results and Discussion}
\label{sec::Results_Discussion}

\subsection{RQ1: Challenges}
\label{rq1_challenges}

\noindent\textbf{Equivalent Mutants (MAJOR):}
During our work on Simulink and Stateflow models, \mutslx produced equivalent mutants.
Those are detected by manual analysis because \mutslx currently does not support automatic detection.
However, we learned that the majority of the equivalent mutants stem from specific operators: Constant Replacement (CR), Relational Operator Replacement (ROR), and, as a special operator, changing the Multiplication parameter of the Product and Gain blocks. 
The latter is representative of the equivalent mutants challenge; hence, we explain it in more detail.

In Simulink, the gain block\footnote{\url{https://nl.mathworks.com/help/simulink/slref/gain.html} (accessed 2026-03-30).} accepts a single input signal and multiplies it by a constant K. The gain can be a real or complex-valued scalar, vector, or matrix.
A typical mutation operator would change multiplication modes from one to another, for example, from element-wise to matrix. However, as long as K is a scalar, mutating multiplication modes makes no difference and provides the same results, regardless of the dimension of the input signal. This causes equivalent mutants.
In the Product block\footnote{\url{https://nl.mathworks.com/help/simulink/slref/product.html} (accessed 2026-03-30).}, however, since multiple signals are multiplied together instead of a constant, equivalent mutants can occur in more scenarios. Changing the multiplication modes makes no difference whether all inputs are scalar or only one is non-scalar. 

The fact that many of the models within \duco contained these blocks made it very difficult to manually perform semantic analysis to confirm that these mutations are equivalent or not.
For these two blocks 
specifically, we implemented a solution that identifies the type (such as real or complex-valued scalar, vector, or matrix)  
of the K constant in the Gain block and the combinations (such as two scalars, a scalar and a vector) of the input signals in the Product block, and applies filtering accordingly. Although this method is not effective for equivalent mutations that change at the block level for gain and product blocks but do not propagate to higher sub-system levels, we are analysing our solution's applicability to equivalent mutations arising in other blocks different from gain and product blocks, since it is successful for mutations that are equivalent directly at the block level.

\noindent\textbf{Requirement Traceability (MAJOR):}
A major challenge we identified during our pilot project is linking mutation results back to requirements.
Organisations maintain their requirements in diverse formats, such as Requirements Toolbox files
(\texttt{.slreqx}), IBM DOORS, Microsoft Word, Microsoft Excel, PDF documents, and other types supported by Simulink.
Each test case is typically linked to one or more requirements, yet mutation testing tools currently do not exploit this traceability information.

\noindent\textbf{Mutation Testing for Stateflow (MINOR):}
Mutation operators for Stateflow transitions and states are primarily implemented using a regular expression–based technique due to limitations in the Stateflow API\footnote{\url{https://nl.mathworks.com/help/stateflow/api/overview-of-the-stateflow-api.html} (accessed 2026-03-30).}, unlike in the Simulink API. This necessity arises because \duco's MATLAB version requires us to use this solution. The Stateflow API in this version does not allow us to modify parameters of transitions and states in Stateflow models. This is a particular solution for a specific MATLAB version implemented in \mutslx. 
Applying these techniques to Stateflow models introduces several challenges.
Different mutation types require specialized regular expressions for numeric, Boolean, logical, and mathematical operations.
Ensuring the syntactically correct placement of mutated expressions is also difficult, particularly when mutations replace operators with alternatives of different string lengths and characters, for example, while replacing ``<'' with ``>='' when the original string is ``<='' results in ``>=='' or remove the logical negation operator (!) for enums (e.g., not in !=) while preserving the structure of the original expression. 
In addition, the expression context must be considered when applying certain mutations.
For example, when mutating the equality operator (==), it is necessary to determine whether the expression belongs to a Boolean or numeric context in order to apply an appropriate mutation (e.g., != or >= ).
Therefore, mutant generation requires not only detecting the operator but also analysing the string as a whole, such as the order of symbols and characters. 
Furthermore, the mutation process must distinguish between expressions located in Stateflow model elements and those appearing in comments. 
Analysing the string as a whole and detecting comments based on a specific identifier (/*...*/) could help resolve those problems. 

\noindent\textbf{Testing \mutslx{} (MINOR):}
During the development of \mutslx, one of the concerns is to design smoke tests to ensure that the \mutslx 
remains functional on a model that includes all types of blocks that existing mutation operators are capable of manipulating. 
These tests verify that the functional integrity is preserved in \mutslx after integration with new features, such as new mutation operators. They are also used to verify the functionality with different MATLAB versions.

\subsection{RQ2: Possible Solutions}
\label{solutions}

This section provides details on the solutions we adopted and the lessons we learned while adopting them.

\noindent\textbf{Tackling Equivalent Mutants:}
We considered several options to address the issue of equivalent mutants.
Based on the literature we first considered Trivial Compiler Equivalence, the dominant technique for code level mutation testing~\cite{papadakis2015trivial}.
As a method for detecting equivalent mutants, the original and mutated code are compiled at the same optimization level to produce two compiled binaries, which are then compared.
If they are identical byte-by-byte, the mutant is marked as equivalent and eliminated.

In the context of Simulink and Stateflow the closest parallel to the binaries are the (\texttt{.slx}) files.
These are binary files, but are in fact compressed XML packages; therefore, comparing the original and mutated model files would not be meaningful they will be the same because there is not compiler optimization performed.
An alternative is to generate C code for both the original and modified models using model-level optimizations with Embedded Coder, then using a compiler such as gcc to perform a second round of optimization on this generated code to produce binary files, and finally compare the results. However, in the practical implementation of this process, several constraints must be taken into account, including the fact that generating code from models is likely to be significantly more expensive than compiling C code.

As an alternative, we took inspiration from using formal verification to show why a mutant is equivalent, or come up with a counterexample as a test case if not~\cite{demeyer2025equivalent}.
As the formal verification we chose the Simulink Design Verifier.
We were also inspired by the back-to-back equivalence testing approach, which verifies that Simulink models and the code generated from them are consistent. This type of test verifies whether the Simulink model and the code generated from it produce the same results as a large input space generated by the Simulink Design Verifier.
The strategy we have implemented works as follows. A broad input space is generated on the original model using Simulink Design Verifier. Subsequently, for each surviving mutant, the original and mutated models are simulated using this input space, and a comparison is made to determine whether they produce the same outputs. If both models produce the same outputs across the entire input space, that mutant is marked as ``likely equivalent'' and eliminated. As a result of this analysis, test cases can be created from the inputs that kill the mutants, thereby amplifying the test suite. 

One drawback we experienced is that this approach generates a very large input space for large models. This results in a large number of simulations and significant time costs, thereby reducing its practical applicability. To make this approach scale, solutions such as filtering the pool of surviving mutants rather than testing all of them could be considered.

\noindent\textbf{Addressing Requirement Traceability:}
Based on feedback from the \duco engineers, we implemented a prototype feature for Signal Builder-based models that leverages Simulink Requirements Toolbox (\texttt{.slreqx}) links.
When a test case (i.e., a Signal Builder group) kills a mutant, the tool traces which requirement is linked to that group, thereby associating killed mutants with the requirements they relate to. This enables identification of which requirements are well-covered by mutation testing
and which may need additional test effort.

In our current tool prototype this traceability support is limited to a Signal Builder and the Simulink Requirements Toolbox.
There are other test execution tools within the Simulink and Stateflow ecosystem and there are many requirement document types used across organizations.
Therefore, for real-world adoption, a customized integration tailored to an organization's specific toolchain is needed.

\noindent\textbf{Handling Mutation Testing for Stateflow:}
To ensure syntactic integrity during operator replacement, our approach leverages regular expressions (regex) that explicitly target the whitespace delimiters surrounding operators. By capturing an operator along with its preceding and succeeding whitespaces (e.g., recognizing < as " < "), the tool safely substitutes it with alternatives of varying lengths, such as " <= " or " > ", completely avoiding malformed constructs. Beyond syntactic correctness, contextual ambiguities—such as whether an equality operator (==) operates in a Boolean or numeric context—are resolved by analyzing the adjacent lexical scope for explicit Boolean literals (e.g., true or false). The expression is treated as logical if such literals are present, and numeric otherwise. Additionally, to avoid mutating code within comments, a pre-processing step delineates comment intervals across the code sequence. By explicitly excluding these blocks from the regex search space, the generation of equivalent or ineffectual mutants is effectively mitigated.

\noindent\textbf{Effective Testing \mutslx{}:}
To achieve this, a diverse set of smoke tests is developed to cover common and representative Simulink and Stateflow use cases while maintaining coverage of existing functionality. To do this, we created several artificial models and extended them as new features were developed, which needed to be tested. This also requires continuously extending test capabilities for newly introduced mutation operators and updating tests when existing operators are refactored. To support this process, a Continuous Integration and Continuous Deployment (CI/CD) pipeline is employed using Jenkins, which automates the execution of tests for both existing and newly implemented features.

\section{Lessons Learned and Practical Insights}
\label{sec::lessons}

This section presents our lessons learned from our pilot project, drawing on the challenges already shared and possible solutions.

\begin{qblock}
\textbf{Lesson 1:} \textit{Equivalent mutants concentrate in a small set of operators, so operator-specific filters outperform general-purpose detection.}
\end{qblock}

When we manually analysed the equivalent mutants generated by \mutslx, we discovered that the majority of them originated from only three operators: Constant Replacement (CR), Relational Operator Replacement (ROR), and mutation of the Multiplication parameter in the Gain/Product blocks.
This led us to develop a dimensionality-tracing filter that works at the block level.
This inexpensive, operator-specific solution eliminated many equivalent mutants without ever introducing them into the simulation, whereas general approaches based on Trivial Compiler Equivalence or Design Verifier induced unacceptable execution times and licensing dependencies for the same mutant pool.

\begin{qblock}
\textbf{Lesson 2:} \textit{Code-level equivalent mutant detection techniques do not transfer to model-level without substantial adaptation.}
\end{qblock}

We attempt to use the Trivial Compiler Equivalence (TCE) method on Simulink and Stateflow models. While (\texttt{.slx}) files are comparable at the byte level, they are actually compressed XML packets, so the same semantics means the same binary assumption on which Trivial Compiler Equivalence is based is not directly applicable. Alternatively, when we tried generating C code with Embedded Coder and comparing the binary after gcc optimisation, we found that code generation was much more expensive than C compilation, that Embedded Coder introduced a license dependency, and that generation was not fully deterministic. We learned that a technique that is straightforward at the code level is not easily applicable at the model level.

\begin{qblock}
\textbf{Lesson 3:} \textit{Requirement traceability in practice is an integration problem; better not consider it a tool feature.}
\end{qblock}

In industry, requirements are stored in diverse formats, such as (\texttt{.slreqx}), IBM DOORS, Microsoft Word, Excel, and PDF. When we developed a prototype traceability feature for the Signal Builder and (\texttt{.slreqx}) combination, even for this single combination, a point to point integration was costly to implement.
More importantly, the solution could not be directly transferred to other test execution tools or other requirement formats. We learned that designing traceability as a single feature integrated into \mutslx was impractical. The more realistic approach is an integration layer that can be customised to the organisation's existing toolchain.
This is why we positioned our (\texttt{.slreqx}) support as a ``proof-of-concept'' rather than a ``general solution''.

\begin{qblock}
\textbf{Lesson 4:} \textit{Fine-tune the mutation operators based
on the bug database of the software team.}
\end{qblock}

Mutation operators for Simulink and Stateflow can be defined and implemented in different layers based on their classes, as shown in Table~\ref{tab:simulink-mutation-classes}. The selection of classes and mutation operators needs to be determined based on company-specific requirements. Then, when an appropriate class is selected, the mutation operators must be customised using the software team's bug database. This makes them more targeted, relevant and impactful. The mutation operators for \mutslx are designed and developed based on this approach, considering the needs and requirements of \duco and fine-tuned using their bug database.


\section{Conclusion}
\label{sec::Conclusion}

Based on shift-left testing principles, several mutation testing tools for Simulink and Stateflow models have been presented.
While these proof-of-concept tools demonstrate the technical feasibility of model-based mutation testing, actual experience reports on how the corresponding tools would work in practice are still lacking.
In this paper, we analyse the challenges we encountered and the lessons we learned when incorporating mutation testing in the context of \duco. 
The main challenges, ranked by severity, are equivalent mutants, requirement traceability, and mutation testing for Stateflow.



\section{Mandatory Data Availability Statement}
\label{sec::Availability}

\small{The authors confirm that the data supporting the findings and experiments of this study are subject to third-party restrictions and cannot be shared due to the high sensitivity of our industrial partner's data. The reader could reach out to the authors for further supporting resources.}



\section{Acknowledgments}
\small{This work is supported by
(a) the Flanders Innovation \& Entrepreneurship (VLAIO) under grant number HBC.2021.0010 entitled ``EFFECTS'';
(b) the Research Foundation Flanders (FWO) under grant number S000323N entitled ``Basecamp Zero'';
(d) Flanders Make, the strategic research centre for the manufacturing industry. (e) the Agency for Innovation \& Entrepreneurship (VLAIO) via the project ``TTRUST'' under grant number HBC20230612. (f) the Industrial Research Fund (IOF) via the project ``MUT4SLX'' under grant number 54104.}

\clearpage
\balance

\bibliographystyle{ACM-Reference-Format}

\bibliography{kavak2026aseBib}

\end{document}